\documentclass[aps,prx,reprint,longbibliography,nofootinbib,superscriptaddress,floatfix]{revtex4-2}

\usepackage{slashed}
\usepackage{verbatim}
\usepackage[T1]{fontenc}
\usepackage{mathbbol}
\usepackage[dvipsnames]{xcolor}
\usepackage{orcidlink}
\usepackage{ragged2e}
\usepackage[normalem]{ulem}
\usepackage[english]{babel}
\usepackage{lipsum}
\usepackage{physics}
\usepackage{dcolumn}
\usepackage{tensor}
\usepackage{comment}
\usepackage{placeins}
\usepackage{graphicx,color,overpic,mathtools}
\usepackage{amsthm,amsmath,amssymb,mathrsfs}
\usepackage{braket,bm,bbm,setspace}
\usepackage{booktabs}
\usepackage{cancel}
\usepackage{float}
\usepackage{xargs}

\usepackage{hyperref}
\hypersetup{
    colorlinks=true,
    linkcolor=RoyalBlue,
    citecolor=ForestGreen,
    urlcolor=RoyalBlue}

\newcommand{\uib}{%
Departament de F\'isica, Universitat de les Illes Balears,
IAC3~--~IEEC, Crta.\ Valldemossa km~7.5, E-07122 Palma, Spain}
\newcommand{\GSSI}{%
Gran Sasso Science Institute (GSSI), I-67100 L'Aquila, Italy}
\newcommand{\GranSasso}{%
INFN, Laboratori Nazionali del Gran Sasso, I-67100 Assergi, Italy}

\begin{document}

\title{Spectral suppression of black hole ringdown tails}

\author{Jose Antonio Le\'on Vega\orcidlink{0009-0002-7382-5137}}
\affiliation{\uib}

\author{Alejandro Svyatkovskyy Kholyavka\orcidlink{0009-0004-1688-6018}}
\affiliation{\uib}

\author{Sayak Datta\orcidlink{0000-0002-4774-0298}}
\affiliation{\GSSI}
\affiliation{\GranSasso}

\author{Xisco Jiménez Forteza\orcidlink{0000-0002-8158-5009}}
\affiliation{\uib}

\begin{abstract}
The late-time power law tail predicted by Price's law is a generic feature of black hole perturbation theory, yet it is largely absent in numerical relativity waveforms of binary black hole mergers. We show that this suppression arises from the spectral structure of oscillatory sources. For a generic perturbation with carrier frequency $\nu$ and characteristic width $\sigma$, the branch-cut excitation coefficient governing the tail is suppressed by $\alpha=\sigma\nu$. For  a Gaussian pulse, the suppression $\sim e^{-\alpha^2/2}$. This suppression is exact and confirmed by the time domain Regge Wheeler evolutions. The same parameter that controls the transition from broadband to frequency selective black hole response is also responsible for the tail suppression.  Moreover, we analytically derive the leading- and next-to-leading-order tail coefficients, finding agreement with numerical fits below the $\sim10\%$ level.  Our results provide a first principle explanation for the absence of tails in quasi-circular mergers and their enhancement in head-on and eccentric ones.
\end{abstract}

\maketitle

\noindent \textbf{\textit{Introduction.}} One of the key puzzles in the black hole (BH) perturbation theory is the strong suppression of late-time tails in numerical relativity (NR) waveforms of quasi circular binary black hole (BBH) mergers while linear black hole perturbation theory (LBHPT) regularly shows existence of it for Gaussian initial pulses, head-on collisions, gravitational collapse and eccentric bound configurations~\cite{Price:1971fb,Leaver:1986gd,Gundlach:1993tp,Ching:1994bd,Ching:1995tj,DeAmicis:2024not,Islam:2024vro}. In this context Price's law is a fundamental prediction of BHPT. Any perturbation of a Schwarzschild BH decays at late-times as a power law $t^{-(2\ell+3)}$ for a time-like observer at fixed $r$, while along future null infinity they decay as $u^{-(\ell+2)}$, where $u={t-r_*}$ denotes the retarded time~\cite{Barack:1999ma,Dafermos:2003yw,Harms:2013ib,Zenginoglu2008}. In rotating spacetimes, the BH response is considerably richer due to mode coupling and the absence of spherical symmetry, and late-time tails have been extensively investigated in~\cite{Hod:1999cu,Barack:1999ma,Barack:1999st,Burko:2007ju,Racz:2011qu,Angelopoulos:2021cpg}.

This stark difference from NR simulations is often explained by claiming that the tail is simply too weak relative to the ringdown, without providing a theoretical explanation. 
Recent perturbative and full NR simulations of eccentric and head on merger, containing enhanced low frequency content, have reported observable late-time tails~\cite{DeAmicis:2024not,DeAmicis:2024eoy,Islam:2024vro,Ma:2024hzq,Cardoso:2024jme}. A theoretical framework predicting nonlinear tails at second perturbative order, together with supporting numerical evidence, has recently been presented in~\cite{Ianniccari:2025avm,Ling:2025wfv,Ling:2026ynd}. The universality of asymptotic late-time tails and their relation to the asymptotic structure of the effective potential is discussed in~\cite{Ching:1995tj,Ching:1995tj,Rosato:2025rtr,Barack:1999wf}. However, despite these recent developments, a fully analytical first-principle understanding of the suppression of late-time tails in generic BBH configurations is  still incomplete. The standard explanation treats the tail suppression as measurement problem rather than a physical one. It does not explain why the amplitude might be small, irrespective of the noise or numerical errors.

Traditionally, existing LBHPT studies that successfully extract tails invariably use non-oscillatory initial data (ID), a pure Gaussian without a carrier frequency~\cite{Ching:1994bd,Ching:1995tj,Andersson:1996cm,Burko:2007ju,Price:2004mm}. However, binary merger sources are fundamentally different. They are oscillatory, with several carrier frequencies set by the orbital dynamics near merger. This systemic distinction has not been exploited.
We show that the effective absence of tails is not specific to NR or nonlinear dynamics, but follows from the spectral structure of the source's oscillations. Oscillatory ID carries a spectral suppression of the branch cut tail coefficient by a single dimensionless parameter $\alpha=\sigma\nu$, where $\sigma$ and $\nu$ are respectively the perturbation's characteristic spatial width and carrier frequency.
The parameter $\alpha$ counts the number of oscillations within the envelope, and controls the suppression factor, determining the observability of the tails.
The key observation is that for localized oscillatory sources, the spectral content of the ID near $\omega\approx 0$ is suppressed, leading to universal suppression of tails.  For Gaussian envelopes this suppression scales as $e^{-\alpha^2/2}$, providing a possible first principle explanation behind absence of tails in waveforms from BBH mergers. 

\begin{figure}[!ht]
    \centering
    \includegraphics[width=\linewidth]{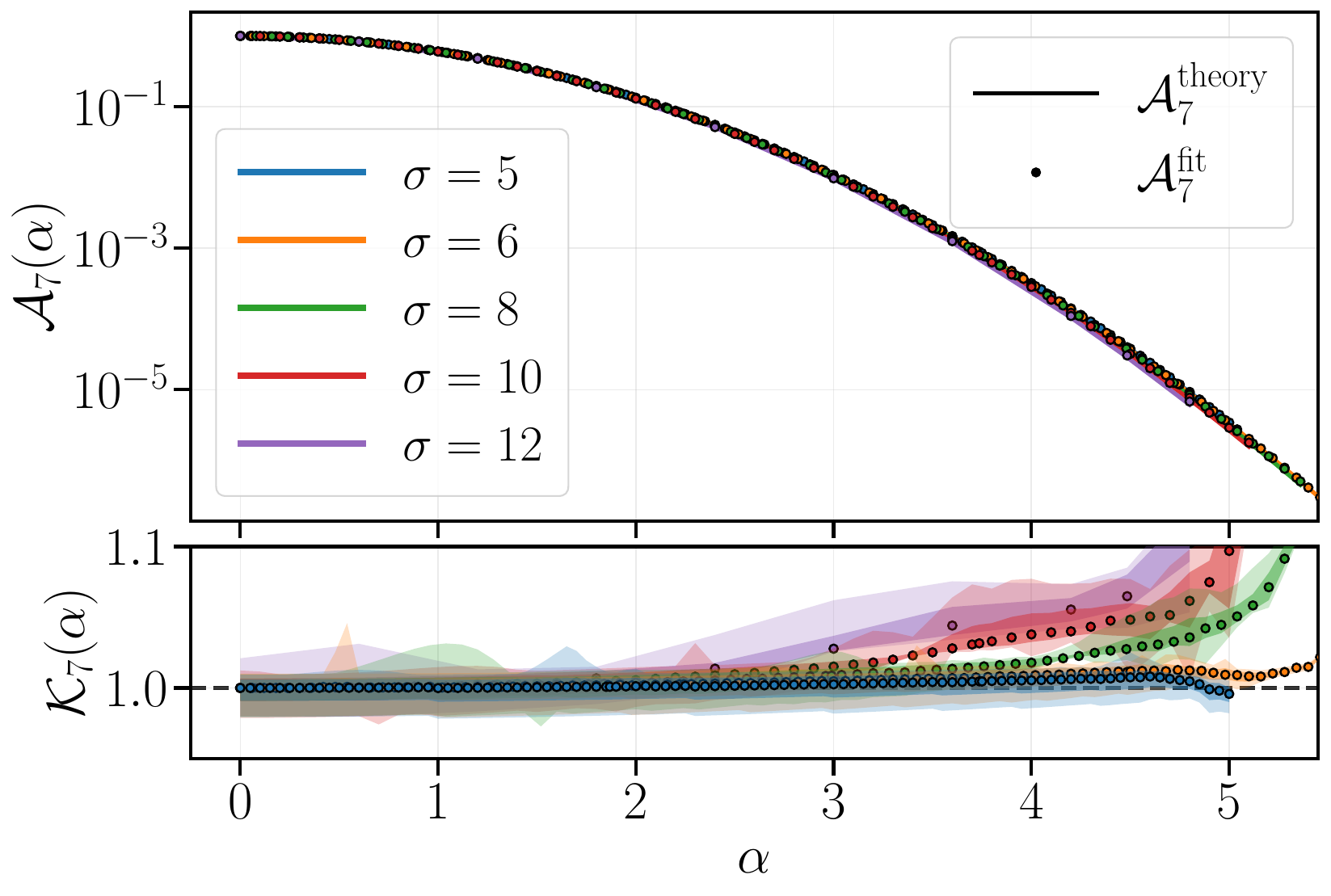}
    \caption{\justifying
    Normalized spectral shape of the leading tail coefficient.
    \textbf{Top:} Normalized coefficient $\mathcal{A}_{7}(\alpha)$ as a function of the dimensionless spectral parameter $\alpha=\sigma\nu$, for several Gaussian widths $\sigma=\{5,6,8,10,12\}$.
    Solid curves show $\mathcal{A}_{7}^{\rm fit}$, while markers denote $\mathcal{A}_{7}^{\rm theory}$.
    \textbf{Bottom:} Normalized ratio $\mathcal{K}_{7}(\alpha)$, quantifying the agreement between the numerical and theoretical spectral shapes.
    Shaded regions denote the $1\sigma$ and $3\sigma$ spread over the accepted fitting windows. }
    \label{fig:a7_norm}
\end{figure}

\noindent \textbf{\textit{Initial data with oscillatory Gaussian.}} To demonstrate the suppression, we consider oscillatory Gaussian ID using the framework developed in \cite{Kholyavka:2026uam},
\begin{equation}
\Psi(r_\star)\Big|_{t=0} = A\,\exp\!\Bigg[-\frac{\big(r_\star - r_0\big)^2}{2\sigma^2}\Bigg]\cos\!\Big[\nu\,\big(r_\star - r_0\big)\Big]\,,
\label{eq:id}
\end{equation}
where $A$ is the amplitude (set to unity), $r_0$ is the pulse center, $\sigma$ is the characteristic spatial width, and $\nu$ is the carrier frequency. 
The initial perturbation is placed at $r_0 \gg 3M$, well outside the photon sphere with ingoing initial condition. 
The spatial Fourier spectrum of~Eq.~\eqref{eq:id} is $\sqrt{2\pi}A\,\sigma\tilde\Psi(\omega) $ with,
\begin{equation}
\tilde\Psi(\omega) =  \,e^{-\frac{\sigma^2}{2}\big(\omega^2 + \nu^2\big)}\,\cosh\big(\sigma\alpha\omega\big)\,.
\label{eq:fourier_id}
\end{equation}
In general, for $\alpha<1$, the spectrum remains concentrated near $\omega\approx0$. As $\alpha$ increases, the spectrum shifts toward $\omega=\pm\nu$, becoming sharply localized there for $\alpha\gg1$. Since $\alpha \simeq 1$ marks the transition from effectively non-oscillatory to spectrally resolved behavior, the $\alpha \gg 1$ regime strongly suppresses the low-frequency support of the perturbation, thus, reducing the tail amplitude.

The late-time tail arises from the branch-cut contribution to the inverse Fourier transform of the Green's function (GF)~\cite{DeWitt:1960fc,Price:1971fb,Leaver:1986gd,Ching:1995tj,Andersson:1996cm, Poisson:2002jz, Casals:2013mpa}. The leading-order contribution follows Price's law~\cite{Price:1971fb},
\begin{equation}
\Psi_{\rm tail}(t)\sim C^{\rm tail}\, t^{-\big(2\ell+3\big)}~~,~~C^{\rm tail} = \tilde{\Psi}(0)\, \mathcal{G}_\ell^{\rm tail}\,.
\end{equation}
where $\mathcal{G}_\ell^{\rm tail}$ depends on both the properties of the spacetime and the source.
Comparing with the quasinormal excitation coefficient (QNEC) $C_n=B_nT_n$~\cite{Andersson:1995zk,Berti:2006wq,berti:2007fi,Zhang:2013ksa,Oshita:2021iyn,Lo:2025njp,DellaRocca:2025zbe,Kubota:2025hjk}, where $B_n$ and $T_n$ are respectively the quasinormal excitation factor (QNEF) and overlap defined in Ref.~\cite{Kholyavka:2026uam}, 
\begin{align}
   \mathcal{Q}_n = \frac{\left|C_n^\mathrm{QNM}\right|}{\left|C^\mathrm{tail}\right|} \propto\left|\frac{B_n}{\mathcal{G}_\ell^{\rm tail}}\omega_n\right| \,\sigma\,e^{\frac{\alpha^2}{2}} e^{\frac{\sigma^2}{2}\big[- (\omega_n^{\rm Re} - \nu)^2 + (\omega_n^{\rm Im})^2 \big]},
   \label{eq:qnm_tail_ratio}
\end{align}
where $n$-th QNM frequency is $\omega_n=\omega^{\rm Re}_n-i\omega^{\rm Im}_n$. At resonance $\nu \approx \omega^{\rm Re}_n$, the ratio of QNM to tail amplitude scales as, $\mathcal{Q}_n \propto  \,\sigma \,{\rm Exp}[\alpha^2/2 + \sigma^2(\omega_n^{\rm Im})^2/2]$ showing that oscillatory sources simultaneously enhance QNM excitation and suppress the tails for $\alpha\gg1$. 

For fundamental mode, at resonance $\nu \approx \omega^{\rm Re}_n \gg |\omega^{\rm Im}_n|$, the exponent is positive and the ratio grows exponentially with $\alpha^2$.  The tail is suppressed relative to the QNM contribution for $\alpha \gg 1$ and remains significant for $\alpha \lesssim 1$, as long as $\mathcal{G}_\ell^{\rm tail}$ does not remove the suppression (as will be shown later).
The exponential suppression $e^{-\alpha^2/2}$ implies that for $\alpha=3$, the tail coefficient is reduced by a factor of $\sim 100$ relative to a non-oscillatory Gaussian of the same width and for $\alpha=5$ by a factor of $10^6$.

\noindent \textbf{\textit{Branch-cut tail suppression.}} 
The response of a BH under perturbation is computed using the system's GF. It has three clearly distinct regimes, early time prompt, intermediate time QNM, and late-time tail. The QNM contribution originates from the pole part of the GF, and the late-time power-law tail comes from the branch-cut contribution. For compact ID, Leaver found the late-time tail in terms of moments of the initial perturbation and velocity as~\cite{Leaver:1986gd},
\begin{equation}
\begin{aligned}
    \Psi_{\rm tail}^{\ell}\big(r_*,t\big)
    &\sim
    \frac{4(-1)^\ell}{\Big[\big(2\ell+1\big)!!\Big]^2}
    r^{\ell+1}
    \Bigg[
    \big(2\ell+3\big)!\, I_{\ell}\big(\psi_0\big)\,\big(t-t_0\big)^{-\big(2\ell+4\big)}
    \\
    &\hspace{0.1cm}
    -
    \big(2\ell+2\big)!\, I_{\ell}\big(v_0\big)\,\big(t-t_0\big)^{-\big(2\ell+3\big)}
    \Bigg].
\end{aligned}
\label{eq:leaver_tail_general}
\end{equation}
where, $I_{\ell}(g)= \int dx\, w_\ell(x)\, g(x)$, and $w_\ell(x)\equiv r(x)^{\ell+1}$, with $x\equiv r_\star$, the Tortoise coordinate.
Thus, the leading timelike late-time behaviour for the $\ell =2$ mode has the structure~\cite{Leaver:1986gd},
\begin{equation}
    \Psi_{\rm tail}^{\ell=2}\big(t,x_{\rm obs}\big)
    \simeq
    a_7\, t^{-7}
    +
    a_8\, t^{-8},
    \label{eq:tail_l2_a7_a8}
\end{equation}
with $a_7 \propto I_2\big(v_0\big)$, $a_8 \propto I_2\big(\psi_0\big)$. In practice, for  numerical fits, $a_7 \approx C^{\rm tail}$.
For  an ingoing ID, $v_0=\partial_x\psi_0$, and $I_{\ell}(v_0)=-\int dx\, w_\ell'(x)\psi_0(x)$, due to vanishing boundary term for localized data~\cite{Kholyavka:2026uam}. Using $A=1$ we find,
\begin{equation}
I_{\ell}\big(v_0\big)=-\sqrt{2\pi}\,\sigma\, r_0^{\ell}\, e^{-\alpha^2/2}\,\mathcal B^{\ell}_v\big(\alpha;r_0,\sigma\big),
    \label{eq:Iv0_B_definition_tail}
\end{equation}
where,
\begin{equation}
\mathcal{B}^{\ell}_v = b^{\ell}_1 + \left(\frac{\sigma}{r_0}\right)^2(1-\alpha^2)\frac{b^{\ell}_3 }{2}+ \left(\frac{\sigma}{r_0}\right)^4(\alpha^4-6\alpha^2+3)\frac{b^{\ell}_5}{24} + \cdots,
\label{eq:Bv_expansion_tail}
\end{equation}
with $r_0^{(\ell+1-n)} b^{\ell}_j\equiv\left. d^j w_\ell/dx^j \right|_{x=r_0}$ . The prefactor $\mathcal{B}^{\ell}_{v}$ varies slowly with $\alpha$ compared to the exponential factor and approaches a constant in the limit $r_0\gg\sigma$. 
The dominant suppression is therefore captured entirely by $e^{-\alpha^2/2}$. Similarly,
\begin{align}
I_{\ell}\big(\psi_0\big)=\sqrt{2\pi}\,\sigma\, r_0^{\ell+1}\,e^{-\alpha^2/2}\,\mathcal B^{\ell}_\psi\big(\alpha;r_0,\sigma\big),
\label{eq:Ipsi0_B_definition_tail}
\end{align}
where,
\begin{equation}
\mathcal{B}^{\ell}_\psi = b^{\ell}_0 + \left(\frac{\sigma}{r_0}\right)^2(1-\alpha^2)\frac{b^{\ell}_2}{2} + \left(\frac{\sigma}{r_0}\right)^4(\alpha^4-6\alpha^2+3)\frac{b^{\ell}_4}{24} + \cdots.
    \label{eq:Bpsi_expansion_tail}
\end{equation}
In the regime with $r_0\gg\sigma$, for \(\ell=2\), $b^{\ell=2}_0=1$ and $b^{\ell=2}_1=3$ giving the leading order term,
\begin{align}
a_7\simeq  \frac{192}{5}\,\sqrt{2\pi}\,\sigma\,r_{\mathrm{obs}}^3 r_0^2e^{-\alpha^2/2},\,a_8\simeq\frac{448}{5}\,\sqrt{2\pi}\,\sigma\,r_{\mathrm{obs}}^3 r_0^3e^{-\alpha^2/2}.
\label{eq:a7_alpha_scaling_tail}
\end{align}
Eq.\eqref{eq:a7_alpha_scaling_tail} shows that the factor $e^{-\alpha^2/2}$ factorizes from the branch-cut moments for both $a_7$ and $a_8$ thus, establishing it as the dominant suppression mechanism.

The details of the derivation and the polynomial prefactor \(\mathcal B^{\ell}_{v,\psi}\) is provided in the Supplemental Material. 
The key point of the above derivation is that the exponential factor $e^{-\alpha^2/2}$ factorizes from the branch-cut moments $I_{\ell}(v_0)$ and $I_{\ell}(\psi_0)$, despite nontrivial radial weight $w_{\ell}(x)$ and derivative structure entering the definition of the tail amplitude. The polynomial prefactors $\mathcal{B}^{\ell}_v$ and $\mathcal{B}^{\ell}_\psi$ encode subleading corrections and vary slowly with $\alpha$. Consequently $\mathcal{G}_\ell^{\rm tail}$ does not remove the suppression as was discussed before. Thus the leading dependence on $\alpha$ is entirely controlled by the universal exponential factor.

The result has a transparent physical interpretation. The branch cut integral receives contributions from the spectral weight of the source near $\omega\approx0$. For a nonoscillatory Gaussian $\alpha\to0$, this weight is maximal and the tail is unsuppressed. For an oscillatory source with $\alpha\gg1$, the ID's Fourier transform is centered at $\pm\nu$, and the weight at $\omega\approx0$ is suppressed, and scaling as $e^{-\alpha^2/2}$ for a Gaussian envelope . The suppression is therefore a direct consequence of spectral peak's location, not of fine tuning.

An important implication of Eq.~\eqref{eq:a7_alpha_scaling_tail} is that tail suppression is not exclusive to NR or nonlinear dynamics; it also emerges within LBHPT for the same class of Gaussian initial data. Keeping the Gaussian envelope fixed while introducing a carrier frequency ($\nu\neq 0$) shifts the spectral weight away from the low-frequency regime ($\omega\approx 0$), thereby suppressing the tail.
In particular, non-oscillatory Gaussian data $(\nu=0)$ thus,  $\alpha=0$, produces the standard unsuppressed Price-law tail. The oscillatory Gaussian data with $\alpha\gg1$ yield an exponentially small tail coefficient despite the GF retaining its branch-cut, and the regime along with being frequency selective also becomes QNM dominating. 
This explains why Gaussian ID studies consistently exhibit tails, while oscillatory merger like sources generically suppress them.

\begin{figure}[t]
\centering
\includegraphics[width=\columnwidth]{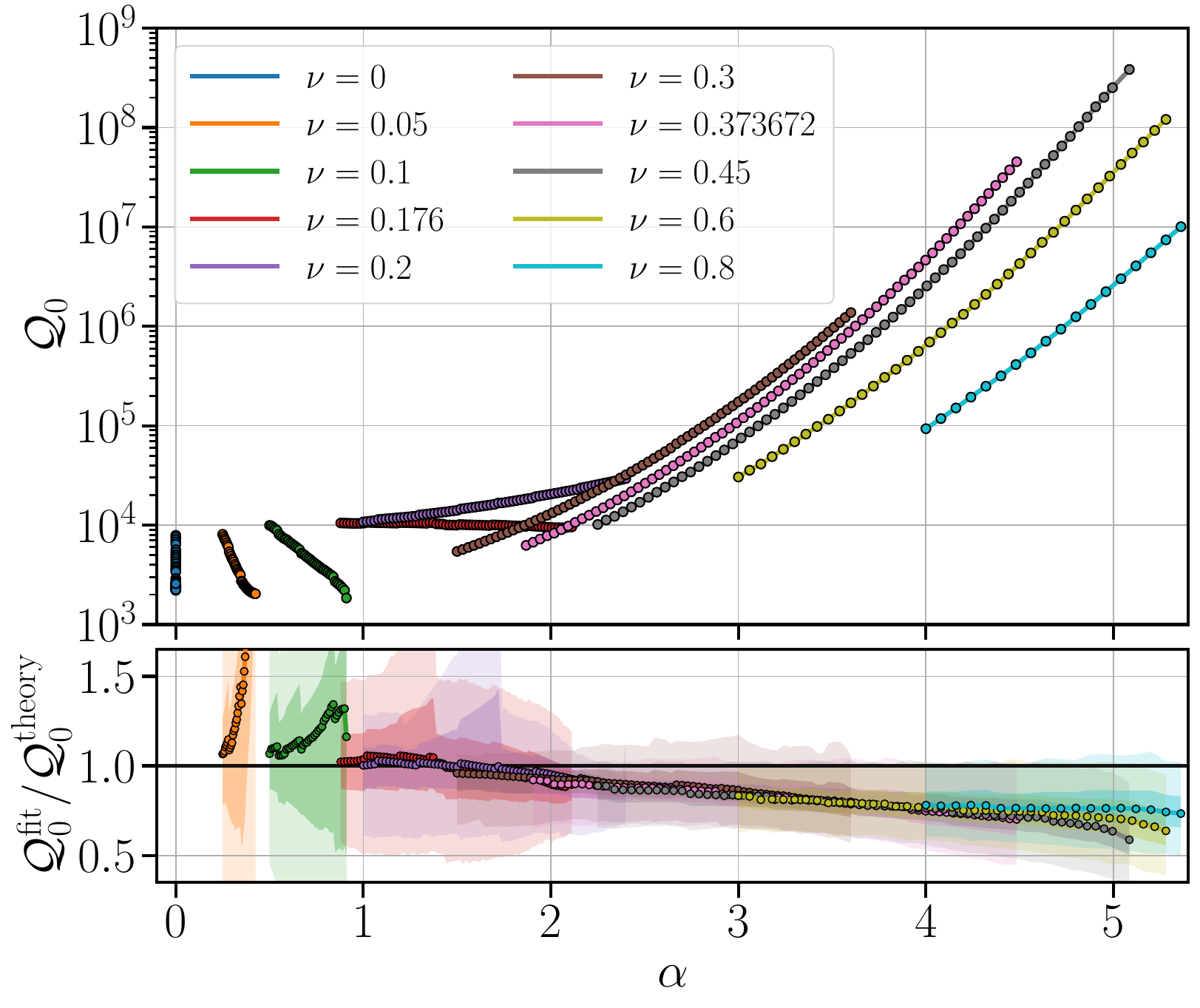}
\caption{\justifying
Ratio of QNM to tail amplitude as a function of $\alpha$ for the fundamental mode $(\ell=2,n=0)$ and $\sigma\in[5,12]$.
The upper panel shows the numerical estimates of $\mathcal{Q}_{0}$ obtained from RW evolutions for several carrier frequencies $\nu$, including the resonant value $\nu=\omega_{0}^{\rm Re}$.
The lower panel shows the ratio between the numerical fit and the full Leaver-based theoretical prediction, $\mathcal{Q}_{0}^{\rm fit}/\mathcal{Q}_{0}^{\rm theory}$.
The overall growth with $\alpha$ follows the expected $e^{\alpha^{2}/2}$ trend, reflecting the simultaneous exponential suppression of the tail and enhancement of QNM excitation.
Shaded bands denote the propagated $1\sigma$ and $3\sigma$ fit uncertainties.} 
\label{fig:ratio}
\end{figure}

\noindent \textbf{\textit{Numerical fitting.}} We confirm our theoretical prediction by numerically solving the Regge--Wheeler (RW) equation and extracting the tail amplitudes $a_7$ and $a_8$.
We extract $a_7$ and $a_8$ by fitting the late-time waveform to Eq.~\eqref{eq:tail_l2_a7_a8} over multiple time windows. We choose $r_0=100$ and the waveforms are extracted at $r_\star^{\rm obs} = 100 $, where the solution is already a good approximation to its asymptotic form~\cite{Kholyavka:2026uam}. The fits are performed in the original evolution time $t$ of the simulation, without shifting the waveform by the arrival time, the peak, or the tail onset. This choice keeps the fitted coefficients in the same time convention as the asymptotic tail expansion.
The tail onset is identified from the waveform-region finder implemented in \texttt{QNMToolkit}~\cite{QNMToolkit2026}, by selecting the region where the local decay is compatible with the expected $\ell=2$ power law, $p=7$. Starting from this onset, we scan multiple windows within the tail-dominated regime and retain only fits with small relative residuals. See the supplemental material for further details. The reported values are the medians over the windows, and the shaded bands show the corresponding $1\sigma$ and $3\sigma$ spread. 

Fig.~\ref{fig:a7_norm} shows the normalized leading tail coefficient $a_7$ extracted from numerical fits, together with the corresponding theoretical prediction, for several Gaussian widths and over the range $\alpha\in[0,5.5]$. For the top panel, we define $\mathcal{A}_{7}^{X}(\alpha)\equiv
\frac{|a_7^{X}(\alpha)|}{|a_7^{X}(0)|}$,
where $X\in\{\mathrm{fit},\mathrm{theory}\}$ denotes the numerical and theoretical results, respectively. The lower panel displays the ratio $
\mathcal{K}_{7}(\alpha)\equiv
\frac{\mathcal{A}_{7}^{\rm fit}(\alpha)}
{\mathcal{A}_{7}^{\rm theory}(\alpha)}$, which quantifies the agreement between the numerical and theoretical spectral values in terms of $\alpha$.  In the top panel, the theoretical prediction and the numerical fit are nearly indistinguishable over the entire $\alpha$ range, exhibiting percent-level agreement and closely following the expected Gaussian suppression $e^{-\alpha^2/2}$. The lower panel quantifies the residual deviation between the numerical and theoretical normalized values, showing that the ratio remains close to unity for all widths considered and $\alpha\lesssim 3$. The shaded bands represent the $1\sigma$ and $3\sigma$ credible region associated with the uncertainty originated from the fits.  With increasing $\sigma$, $\mathcal{K}_{7}(\alpha)$ mildly deviates from one and increases with $\alpha$, induced by the terms $\mathcal{O}(\alpha^2)$ and terms $\mathcal{O}(\sigma^2/{r_0^2})$ in Eq.\eqref{eq:Bpsi_expansion_tail}, together with the identification challenges of the optimal fitting window.

In the top panel of Fig.~\ref{fig:ratio}, we show the ratio $\mathcal{Q}_0$ between the fundamental $n=0$ QNM and the tail amplitude in terms of $\alpha$, for  $\sigma\in[5,12]$. The ratio is found by fitting the QNM and tail amplitudes from the time domain RW evolutions using the algorithm implemented in~\cite{Kholyavka:2026uam}. To evaluate $\mathcal{Q}_0$, the tail amplitude is estimated around the tail onset time, around $t\gtrsim 100M$  after the waveform peak amplitude, for each simulation with different $\nu$ and $\sigma$.
For $\alpha=\nu=0$, the different points represent different $\sigma$ and they all lie on the same vertical line. For $\nu=0.05,\,0.1$, $\mathcal{Q}_0$ decays with $\alpha$ and the  tail contribution grows compared to QNM, which is consistent with the results obtained in~\cite{Kholyavka:2026uam}. At critical carrier frequency $\nu_c\simeq0.176$, the slope of $\mathcal{Q}_0$ changes sign, marking the transition between tail- and QNM-dominated behavior. This critical frequency is predicted by Eq.~\eqref{eq:qnm_tail_ratio} by requiring the exponent $\nu^2-(\omega_n^{\rm Re}-\nu)^2+(\omega_n^{\rm Im})^2$ to vanish, yielding $\nu_c=[(\omega_n^{\rm Re})^2-(\omega_n^{\rm Im})^2]/(2\omega_n^{\rm Re})\simeq0.176$, in agreement with the numerical result. For $\nu>\nu_c$, the ratio between the $n=0$ QNM and the tail amplitude grows with increasing $\alpha$, reflecting the exponential suppression of the tail predicted by Eqs.~\eqref{eq:qnm_tail_ratio}, \eqref{eq:Iv0_B_definition_tail}, and \eqref{eq:Ipsi0_B_definition_tail}. In the bottom panel, we show the ratio $\mathcal{Q}_0^{\rm fit}/\mathcal{Q}_0^{\rm theory}$, with $\mathcal{Q}_0^{\rm theory}$  defined in Eq.~\eqref{eq:qnm_tail_ratio}. The shaded bands correspond to the $1\sigma$ and $3\sigma$ uncertainty regions obtained from the  numerical fits. Notice that, for $\mathcal{Q}_0^{\rm fit}$, the fractional deviations can become larger due to the propagation of uncertainties  sourced by the tail and QNM fits. This effect is particularly pronounced at low $\alpha$, where the fitted $n=0$ QNM amplitude is more sensitive to the choice of fitting window~\cite{Kholyavka:2026uam}. However, despite the larger fractional deviations, the fitted bands remain consistent with unity throughout the entire range explored. 

\noindent \textbf{\textit{Physical implications.}} 
\label{sec:Physical_implications}
The present results establish the suppression mechanism and its parametric dependence within LBHPT for the prescribed ID. A direct quantitative connection to BBH merger sources requires identifying the effective $(\sigma_{\rm eff},\nu_{\rm eff},\alpha_{\rm eff})$ for the post-merger perturbation from the binary's orbital history, which is beyond the scope of the present work. Nevertheless, the spectral picture is qualitatively consistent with existing observations. 
Quasi-circular mergers produce oscillatory perturbations concentrated near the QNM frequency, suggesting $\alpha_{\rm eff}\gg1$ and strong tail suppression. 
Eccentric mergers naturally provide the setting for this mechanism to become observable. Unlike in circular binaries, they show burst-like dynamics and also contain richer harmonics structure~\cite{Barack:2003fp, Drasco:2003ky, Drasco:2005kz, Babak:2006uv, Yunes:2009yz,Gold:2012tk,Huerta:2019oxn, Hughes:2021exa, Islam:2024zqo}, providing low frequency support. Under the current framework, this corresponds to a reduced  $\alpha_{\rm eff}$ and therefore an enhanced tail amplitude. Such behavior is consistent with the tails recently uncovered in fully nonlinear eccentric simulations \cite{DeAmicis:2024eoy,DeAmicis:2024not,Islam:2024vro}.
As a plausible estimate, for a quasi-circular BBH merger, the GW frequency at the $(\ell m)=(22)$ mode amplitude peak is $\omega_{\rm peak}\sim 0.28-0.45$ \cite{buonanno:2006ui, pan:2011gk, Healy:2017mvh, Healy:2018swt}, giving $\alpha_{\rm eff} = [0.28,0.45]\sigma_{\rm eff}$. For such high frequencies, $\alpha_{\rm eff}\gtrsim 1$ for $\sigma_{\rm eff} \gtrsim [2.2,3.5]$, suggesting the tail-suppression regime is easily reached in full NR simulations. A precise identification of the effective parameters from binary parameters are left for the future.

The suppression mechanism identified here is not tied specifically to Gaussian envelopes. More generally, for localized oscillatory perturbations of the form $\Psi(u)|_{t=0}=E(u)\cos(\alpha u)$, the branch-cut excitation is controlled by the spectral weight of the envelope near $\omega\approx0$. In particular, the tail amplitude at leading order and in the far field zone, inherits the Fourier decay properties of the envelope itself. Consequently, oscillatory sources suppress tails whenever increasing $\alpha$ depletes low-frequency support, while the precise suppression law depends on the envelope profile. The generality of this mechanism is discussed in the Supplemental Material. In the current work we used Gaussian profile as a convenient analytical model, but the suppression mechanism is more general. Any source whose radially weighted Fourier transform is concentrated away from $\omega\approx 0$ will suppress the branch-cut integral.  The Gaussian profile provides us the canonical example.

\noindent \textbf{\textit{Conclusion.}} We have identified a spectral mechanism for the suppression of Price-law tails in LBHPT using the framework developed in Ref.~\cite{Kholyavka:2026uam}. For oscillatory ID with dimensionless parameter $\alpha=\sigma\nu$, the branch-cut tail coefficient is suppressed as $e^{-\alpha^2/2}$, exact in the asymptotic limit and validated against time domain RW evolution. The suppression is a direct consequence of the location of spectral power in frequency space.
It does not require nonlinear evolution. Even traditional Gaussian class ID can suppress tails under LBHPT. The oscillatory sources concentrate their Fourier spectrum away from $\omega\approx 0$, where the branch cut contribution to the late-time tail is generated. The parameter $\alpha$ plays a dual role. It simultaneously controls the onset of the resonant QNM excitation and the exponential suppression of the tail.
Whether this qualitative picture survives a quantitative mapping from orbital parameters in BBH merger to $\alpha$ remains an open and observationally relevant question.

If such a mapping exists, $\alpha$ may provide an observable characterization of merger spectral structure complementary to QNM frequencies. In this picture, the post-merger response is determined not only by the spacetime spectrum itself, but also by how source dynamics distribute spectral weight across the structures of the Green function.
While the present analysis has focused on Schwarzschild BHs, the suppression mechanism is fundamentally spectral and therefore expected to extend to rotating spacetimes. In particular, studies of tail behavior in Kerr spacetimes, which exhibit qualitatively similar features to those found in Schwarzschild spacetimes, are reported in~\cite{Hod:1999cu,Barack:1999ma,Barack:1999st,Burko:2007ju,Racz:2011qu,Angelopoulos:2021cpg}. Since branch-cut excitation is controlled by the low-frequency content of the perturbation, oscillatory sources in Kerr are likewise expected to suppress the tail contribution, which provides an interesting avenue for future investigation.

\begin{acknowledgments}
%
This work was supported by the Universitat de les Illes Balears (UIB)
with funds from the Programa de Foment de la Recerca i la Innovaci\'o
de la UIB 2024-2026 (supported by the yearly plan of the Tourist Stay
Tax ITS2023-086); the Spanish Agencia Estatal de Investigaci\'on grants
PID2022-138626NB-I00, RED2024-153978-E, RED2024-153735-E, funded by
MICIU/AEI/10.13039/501100011033 and the ERDF/EU; and the Comunitat
Aut\`onoma de les Illes Balears through the Conselleria d'Educaci\'o i
Universitats with funds from the ERDF (SINCO2022/18146 - Plataforma
HiTech-IAC3-BIO).
X.J.\ is supported by the Spanish Ministerio de Ciencia, Innovaci\'on
y Universidades (Beatriz Galindo, BG22-00034) and cofinanced by UIB.
S.D.\ acknowledges financial support from MUR, PNRR - Missione~4 -
Componente~2 - Investimento~1.2 - finanziato dall'Unione europea -
NextGenerationEU (cod.\ id.:\ SOE2024\_0000167, CUP:
D13C25000660001).
The authors thankfully acknowledge the computer resources at
MareNostrum~5 and the technical support provided by the Barcelona
Supercomputing Center (BSC) through grants No.~RES-AECT-2025-1-0011,
RES-AECT-2025-2-0038, and RES-AECT-2025-3-0050 from the Red
Espa\~nola de Supercomputaci\'on (RES).
The authors acknowledge CINECA for providing high-performance computing resources and support through the ISCRA initiative under project HP10CU7X29.

\end{acknowledgments}

\bibliography{main}

\clearpage
\section*{Supplemental Material}
\section{Derivation of exponential suppression}
\label{app:weight_expansion}
We define $y=x-r_0$ and $u=y/\sigma$ to express the ID from Eq.~\eqref{eq:id} as,
\begin{equation}
\Psi(u)|_{t=0} = A e^{-u^2/2}\cos(\alpha u).
\end{equation}
The Leaver moment $I(v_0)$ involves integrals like
\begin{align}
    I_{\ell}(v_0) =& -A\sigma\int_{-\infty}^{+\infty}  \partial_x w_{\ell} e^{-u^2/2} \cos(\alpha u)\, du \notag\\[0.5em]
    =& -A\sigma\sum_{n=0}^{\infty} \frac{(\sigma)^n}{n!} \bar{b}^{\ell}_{n+1}(r_0)J_n 
\end{align}
where $w_\ell(x) = r(x)^{\ell+1}$ and,
\begin{align}
    \partial_x w_{\ell} =& \sum_{n=0}^{\infty} \frac{(\sigma u)^n}{n!} \Bar{b}^{\ell}_{n+1}(r_0)\\[0.5em]
    J_n=& \int_{-\infty}^{+\infty}  u^n e^{-u^2/2} \cos(\alpha u)\, du,
\end{align}
with, $\bar{b}^{\ell}_n = w_{\ell}^{(n)}(r_0)$ where $n$ in $w_{\ell}$ parentheses denotes $n$-th derivative with respect to $x$.
$J_n$ can be generated from $\int du e^{-u^2/2}e^{i\alpha u}=\sqrt{2\pi}e^{-\alpha^2/2}$ as,
\begin{align}
    J_n=&{\rm Re}\left[(i)^n(-\partial_{\alpha})^n\sqrt{2\pi}e^{-\alpha^2/2}\right]\nonumber\\[0.5em]
    =& \sqrt{2\pi}\,e^{-\alpha^2/2}\Re\!\left[(i)^n {\rm He}_n(\alpha)\right],
\end{align}
where ${\rm He}_n$ is the $n$-th Hermite polynomial. Defining $b^{\ell}_n
\equiv\, \bar b^{\ell}_nr_0^{(n-\ell-1)}$ we get,
\begin{equation}
I_{\ell}(v_0) = -A\sqrt{2\pi}\sigma r_0^{\ell}\, e^{-\alpha^2/2} \mathcal{B}^{\ell}_v(\alpha;r_0,\sigma),
\end{equation}
with the polynomial
\begin{equation}
\mathcal{B}^{\ell}_v = b^{\ell}_1 +  \left(\frac{\sigma}{r_0}\right)^2(1-\alpha^2)\frac{b^{\ell}_3 }{2}
+ \left(\frac{\sigma}{r_0}\right)^4(\alpha^4-6\alpha^2+3)\frac{b^{\ell}_5}{24} + \cdots
\end{equation}
from Eq.~\eqref{eq:Bv_expansion_tail}, and $b^{\ell}_n$ depends on $r_0$ as $2M/r_0$.
In the limit $\sigma/r_0 \ll 1$, the prefactor becomes $\mathcal{B}^{\ell}_v \sim b^{\ell}_1=(\ell+1)$, and the corrections scale as $(\sigma/r_0)^{2n}$. So
\begin{equation}
I_{\ell}(v_0) \approx -A\sqrt{2\pi}\sigma(\ell+1) \, r_0^\ell \, e^{-\alpha^2/2}.
\end{equation}
This justifies the far-field scaling in Eq.~\eqref{eq:a7_alpha_scaling_tail}.
Derivation of $I_{\ell}(\psi_0)$ is identical except $\partial_x w_{\ell}(x)\to w_{\ell}(x)$. From Eq.~\eqref{eq:tail_l2_a7_a8} we find,
\begin{align}
    I_{\ell}(\psi_0) =& A\sigma\sum_{n=0}^{\infty} \frac{(\sigma)^n}{n!} \bar{b}^{\ell}_{n}(r_0)J_n \nonumber\\[0.5em]
=& A\sqrt{2\pi}\sigma r_0^{\ell+1}\, e^{-\alpha^2/2} \mathcal{B}^{\ell}_\psi(\alpha;r_0,\sigma),
\end{align}
which is Eq.~\eqref{eq:Ipsi0_B_definition_tail}, with
\begin{equation}
\mathcal{B}^{\ell}_\psi = b^{\ell}_0 + \left(\frac{\sigma}{r_0}\right)^2(1-\alpha^2)\frac{b^{\ell}_2}{2} 
+ \left(\frac{\sigma}{r_0}\right)^4(\alpha^4-6\alpha^2+3)\frac{b^{\ell}_4}{24} + \cdots
\end{equation}
from Eq.~\eqref{eq:Bpsi_expansion_tail}. In the far-field limit $\sigma/r_0 \ll 1$, the prefactor becomes $\mathcal{B}^{\ell}_{\psi}\sim b^{\ell}_0=1$, with the corrections scaling as $(\sigma/r_0)^{2n}$. So
\begin{equation}
I_{\ell}(\psi_0) \approx A\sqrt{2\pi}\sigma \, r_0^{\ell+1} \, e^{-\alpha^2/2}.
\end{equation}
In the far zone limit $2M/r_0\to0$ we find,
\begin{align}
b^{\ell}_n=&\frac{(\ell+1)!}{(\ell+1-n)!},\qquad n\le \ell+1,\\[0.5em]
B^{\ell}_v=&(\ell+1)\sum_{n=0}^{\lfloor\ell/2\rfloor}
\frac{\ell!}{(\ell-2n)!}\frac{(-1)^n}{(2n)!}
\left(\frac{\sigma}{r_0}\right)^{2n}{\rm He}_{2n}(\alpha),\\[0.5em]
B^{\ell}_\psi=&\sum_{n=0}^{\lfloor(\ell+1)/2\rfloor}
\frac{(\ell+1)!}{(\ell+1-2n)!}\frac{(-1)^n}{(2n)!}
\left(\frac{\sigma}{r_0}\right)^{2n}
{\rm He}_{2n}(\alpha).
\end{align}

\section{Generality of the suppression}

The exponential suppression factor $e^{-\alpha^2/2}$ derived in the main text and in Sec.~\ref{app:weight_expansion} of supplemental material is a consequence of Gaussian envelope, whose Fourier transformation is also a Gaussian. In this section we demonstrate that the suppression mechanism is more general. We show that the functional form of the suppression is determined by ID's envelope, and any sufficiently localized oscillatory source can produce a suppressed tail, while the nature of the suppression is determined by the envelope's Fourier amplitude at $\omega\approx 0$, $\tilde{\Psi}(0)$. For a generic real oscillatory ID, with a sufficient localized envelope which satisfies $
\left(\frac{\sigma}{r_0}\right)^n
\left|
\frac{J_n(\alpha)}{J_0(\alpha)}
\right|
\ll 1,\,
\,\text{with}\, n\geq 1,
$ of the form 
\begin{align}
    \Big.\Psi(u)\Big|_{t=0}=E(u)\cos(\alpha u),\quad u=(x-r_0)/\sigma,
\end{align}
The spectral weight near zero frequency is,
\begin{align}
    \tilde{\Psi}(0;\alpha)\propto \int_{-\infty}^{\infty}E(u)\cos(\alpha u) du,
\end{align}
In the regime,
$\sigma\ll r_0$, the radial weights vary slowly across the support of the packet. At
leading order it can be approximated by $w_\ell(r_0+\sigma u)\approx w_\ell(r_0)+\mathcal{O}(\frac{\sigma^2}{r_0^2})$ and equivalently for $w'_{\ell}(r_0+\sigma u)\approx w'_{\ell}(r_0) + \mathcal{O}(\frac{\sigma^2}{r_0^2})$. To see why the radial-weight corrections are subleading, recall that $w_\ell(x)=r(x)^{\ell+1}$, and in the far field \(r(x)\sim x\), and therefore
\begin{equation}
w_\ell(x)\sim x^{\ell+1}\,.
\end{equation}
For \(\ell=2\), the leading far-zone behaviour is \(w_2(x)\sim x^3\). Hence derivatives above third
order vanish. For example, at \(x=100\), \(r\simeq92.38\) and
$w_2'''(100)\simeq5.87,
    \,\,
    w_2^{(4)}(100)\sim10^{-3},
    \,\,
    w_2^{(5)}(100)\sim- 10^{-5}$.

The dominant dependence on the oscillation parameter \(\alpha\) is therefore controlled
by the envelope factor $\tilde{\Psi}(0;\alpha)$, while the radial corrections only modify
the slowly varying prefactor.
Hence the
leading tail amplitude inherits the same envelope factor and different envelope choices yield different suppression laws,
\begin{align}
    \Psi_{\rm tail}
    \sim
    \tilde{\Psi}(0;\alpha)\,t^{-(2\ell+3)},
\end{align}
up to radial prefactors and subleading inverse-time corrections.
This demonstrates that the suppression mechanism identified in the main text is not unique to Gaussian ID. 
More generally, the suppression mechanism only requires the relevant moments, $I_{\ell}(v_0)$ and $I_{\ell}(\psi_0)$, of the localized oscillatory source to be well defined and to decrease as \(\alpha\) increases. The Gaussian provides the strongest canonical case with $\tilde{\Psi}(0;\alpha)\propto\,e^{-\alpha^2/2}$.

\section{Numerical methods and Suppression of $a_8$ tail}
\label{app:results_a8}
In this section we describe the fitting method used to construct the plots. The extraction of the tail coefficients is performed by subtracting the theoretical contribution of the neighbouring order before fitting the desired coefficient. For the leading coefficient we use
\begin{equation}
t^7\left[\Psi(t)-a_8^{\rm theory}t^{-8}\right]=a_7^{\rm fit}+\mathcal{O}(t^{-2}),
\end{equation}
while for the subleading coefficient we use
\begin{equation}
t^8\left[\Psi(t)-a_7^{\rm theory}t^{-7}\right]=a_8^{\rm fit}+\mathcal{O}(t^{-1}).
\end{equation}
The fit is repeated over the accepted late-time windows, and the final quoted value is the median over this ensemble as in~\cite{Kholyavka:2026uam}. 

\begin{figure}[ht!]
    \centering  \includegraphics[width=\linewidth]{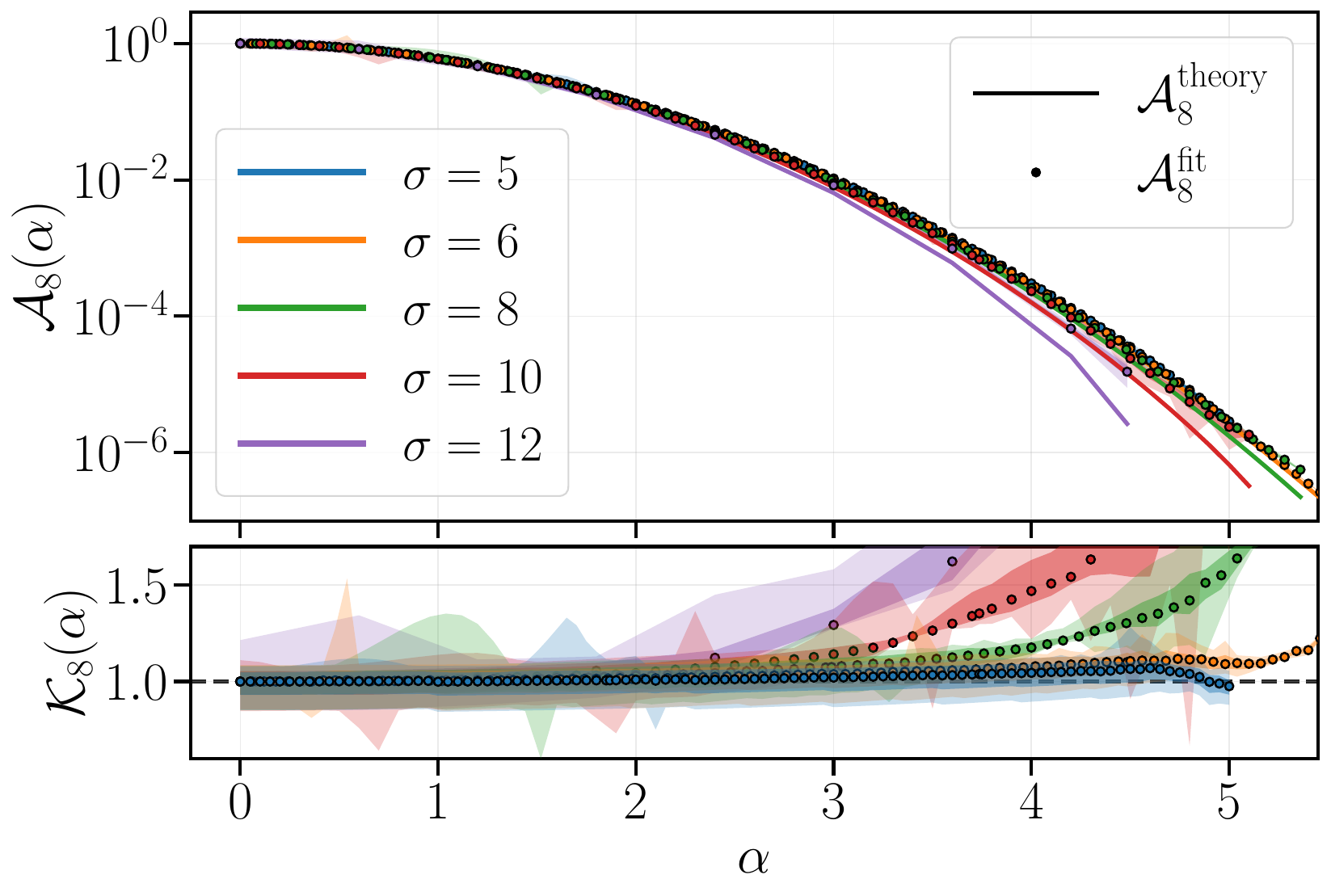}
    \caption{\justifying 
    Same analysis as in Fig. \ref{fig:a7_norm}, but for the subleading tail coefficient $a_8$.
    \textbf{Top:} Absolute value of the cleanly extracted coefficient $\mathcal{A}_8^{\rm fit}$ as a function of the dimensionless spectral parameter $\alpha$, compared with the full theoretical prediction $\mathcal{A}_8^{\rm theory}$ for several Gaussian widths $\sigma=\{5,6,8,10,12\}$.
    \textbf{Bottom:} Normalized ratio $\mathcal{K}_8(\alpha)$, quantifying the agreement between the numerical extraction and the full theoretical result. Shaded regions denote the $1\sigma$ and $3\sigma$ spread over the accepted fitting windows.}
    \label{fig:a8}
\end{figure}

In Fig.~\ref{fig:a8} we show $\mathcal{A}_8(\alpha)$ (top panel) and $\mathcal{K}_8(\alpha)$ (bottom panel) resulting from our numerical fits, and from the theoretical value shown in Eq.~\eqref{eq:a7_alpha_scaling_tail}. We observe good agreement between the theoretical and numerical values at low $\alpha$. Remarkably, the fitted $\mathcal{A}_8(\alpha)$ reproduces the expected exponential suppression predicted by the theory. For $\alpha \gtrsim 3$, the agreement mildly deteriorates. In particular, $a_8$ becomes increasingly subdominant, making its extraction and the identification of optimal fitting window more challenging. 

\end{document}